# ANOMALOUS PHASE DIAGRAM IN SIMPLEST PLASMA MODEL

Igor L. Iosilevski and Alexander Yu. Chigvintsev

Moscow Institute of Physics and Technology, Dolgoprudny 141700, Russia

## INTRODUCTION

Problem of Phase Transition (PT) in Strongly Coupled Coulomb Systems (SCCS) is of great interest in plasma theory during very long time[1-5]. Besides the study of hypothetical PT in real plasmas[1-3] a complementary approach is developing[4,5] where the main subject of interest is definitely existing PT in simplified plasma models. In our previous study[6-8] we dealt with a phase transition in the set of plasma models with common feature - combination of (i) absence of individual correlations (coupling) between charges of opposite sign, and (ii) total compressibility of system. The simplest example of such a system is One Component Plasma (OCP) on uniform, but *compressible* compensating background (following notation - OCP{c}). The well-known *prototype* model is OCP with a *rigid* background (notation - OCP{r}). This variant of OCP is studied carefully nowadays[9,10]. It can not collapse or explode spontaneously. The only phase transition - crystallisation - occurs in OCP{r} without any density change.

Transition to the OCP on uniform and *compressible* background leads to appearance of a *new* first-order *phase transition* of gas-liquid type[6]. New phase diagram combines previous crystallisation, now with a *finite* density change, with a qualitatively different coexistence curve of the new phase transition. The structure and parameters of this phase diagram strongly depend on exact definition of thermodynamic contribution of background. The simplest variant of OCP{r} is the «Single OCP» - the system of classical point charges with a compressible background of ideal fermi-gas of electrons. This variant of OCP was declared repeatedly[11,12] but the discussed phase transition was out of consideration. Closely similar structure of global phase diagram was obtained in «Combined OCP»[6-8]. This is superposition of two *non-coupled* OCP-s of mass-non-symmetrical charged particles of opposite sign.

## PHASE DIAGRAM OF SINGLE OCP{c}

Three qualitatively different situations should be distinguished for the OCP{c} depending on the value of charge number $Z$:

1) Low value of charge number - $Z < Z_1^* \approx 35$
2) High value of charge number - $Z > Z_2^* \approx 45$
3) Intermediate value of charge number - $Z_1^* < Z < Z_2^*$

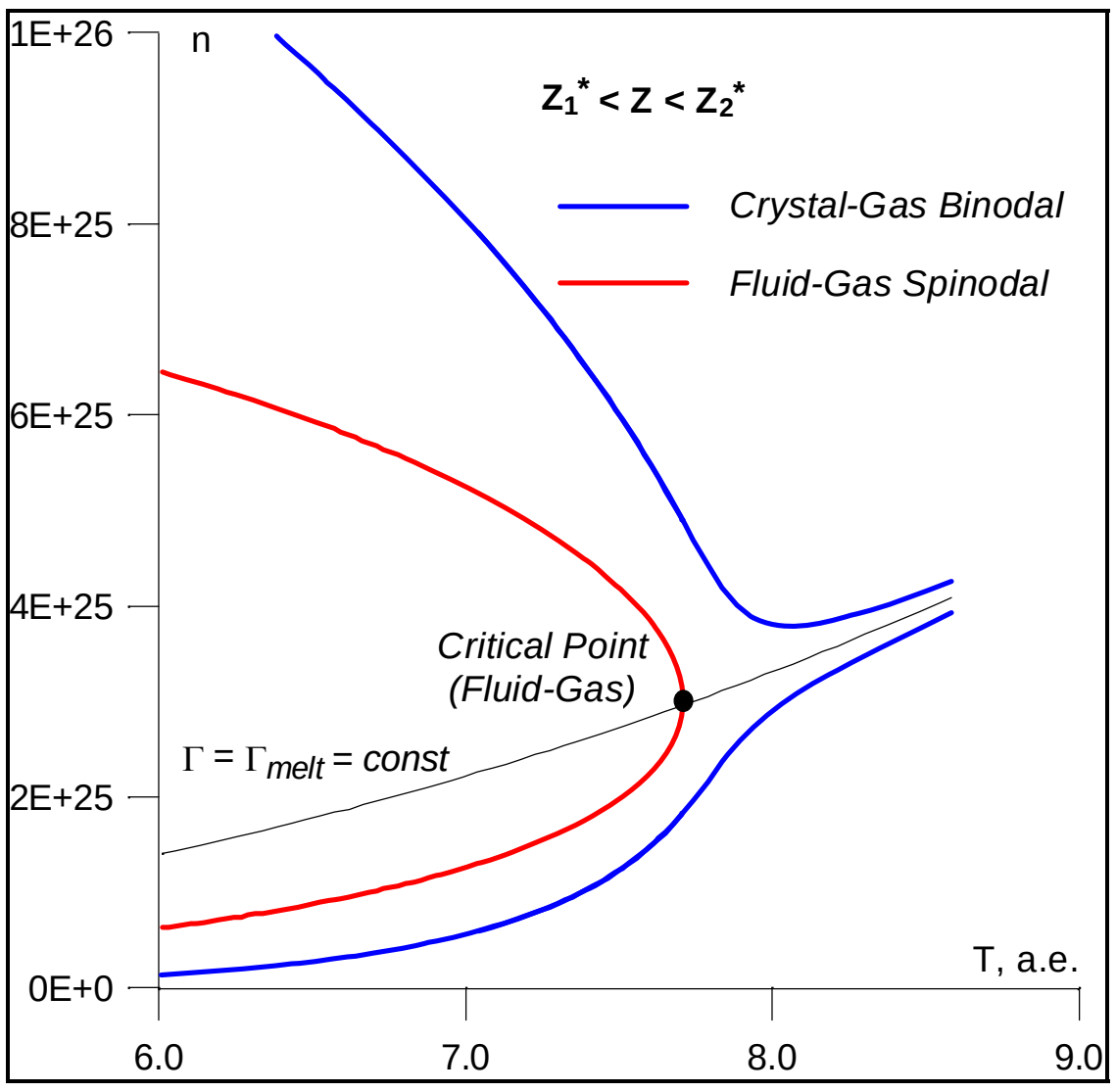

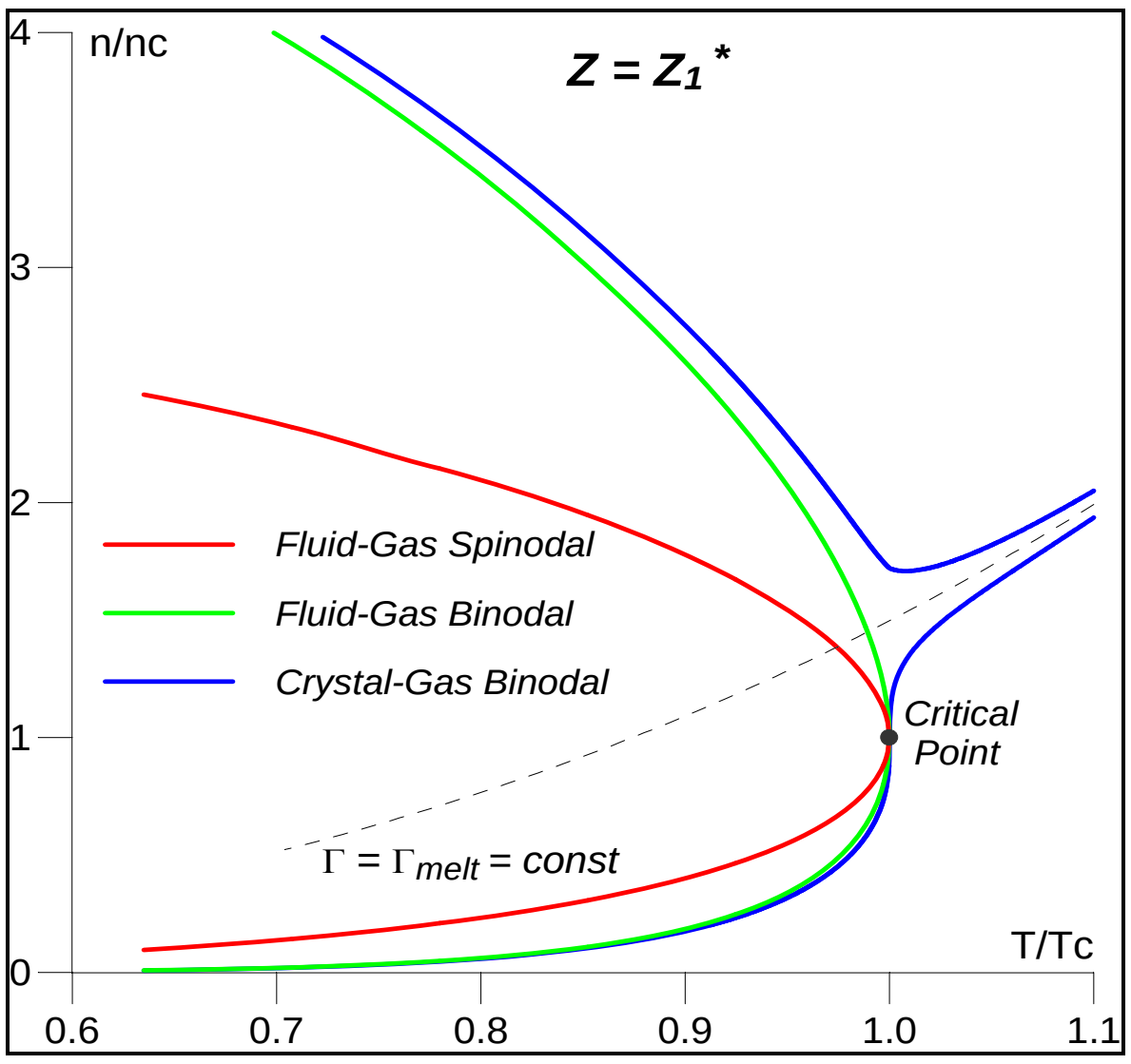

**Figure 1.** Phase diagram of the Single OCP{c} of classical point charges on uniform compressible background of ideal fermi-gas of electrons at intermediate value of charge number ($Z_1^* < Z = 40 < Z_2^*$). Notations: *1*- melting line of prototype OCP{r} ($\Gamma \sim \Gamma_{melt} \approx 178$); *2′,2″,3′,3″*- global crystal-fluid (gas and liquid) coexistence: *2′,2″* - melting, *3′,3″*- sublimation; *4* - spinodal of *metastable* liquid-gas coexistence; *5* - its critical point (*cp*).

**Figure 2.** The same for the lowest of two boundary values of charge number ($Z = Z_1^* \approx 34.6$). *1-4* - as at Figure 1; *5* - pseudo-critical «termination» point; *6′,6″* - *metastable* liquid-gas binodal.

## Low Values of Charge Number ($Z \sim 1$)

Phase diagram of the model was carefully studied in[6-8]. The *ordinary* structure of global phase diagram was obtained in this case: the relative position of critical and triple points, melting «stripe», gas-liquid and gas-crystal coexistence, are totally equivalent to those for normal substances.

## High Values of Charge Number ($Z \sim 100$).

Highly anomalous structure of global phase diagram was announced at previous study[6,7]. The melting «stripe» ($\Gamma \approx 178$) crosses *gaseous* part of coexistence curve of the new phase transition.

- Triple point is placed at *gaseous* part of global phase boundary.
- Critical point is placed at *crystalline* part of global phase boundary.
- Crystal-crystal coexistence of two *dense* and *expanded* crystalline phases of the *same structure* occurs in OCP{r} at such a high values of charge number *Z*.

## Intermediate Values of Charge Number ($Z_1^* < Z < Z_2^*$)

The most remarkable *anomalous* phase diagram corresponds to the case when the melting line of prototype OCP{r} ($\Gamma \sim \Gamma_{melt} \approx 178$) crosses coexistence curve of the new gas-liquid phase transition just closely to its *critical point.* As a result of this coincidence:

- The *only* phase transition exists in the model. It corresponds to the *global crystal - fluid coexistence* – continuous superposition of melting and sublimation (see Figure 1).
- There is *no* true critical point.
- There is *no* triple point.

- Coexistence curve in $P \leftrightarrow T$ (pressure$\leftrightarrow$temperature) plane is a *continuous, infinite* curve. There is *no* any break at this curve.

**Boundary Values of Intermediate Charge Number Interval** ($Z_1^* < Z < Z_2^*$)

Remarkable feature of phase diagram of OCP{c} at $Z = Z_1^*$ or $Z = Z_2^*$ is an existence of *pseudo-critical point* where the well-known standard conditions are fulfilled:

$$(\partial P/\partial V)_T = 0 \qquad (\partial^2 P/\partial V^2)_T = 0$$

$Z = Z_1^* \approx 34.6^*$ – on *gaseous part* of crystal-fluid binodal (see Figure 2)
$Z = Z_2^* \approx 45.4^*$ – on *crystalline part* of crystal-fluid binodal.

When we use the same as in[6-8] analytical fits for equation of state of both subsystems, OCP{r} and background, we obtain following parameters of the both pseudo-critical points:

**Table 1.** Parameters of pseudo-critical point in OCP of classical point charges on the uniform and compressible background of ideal fermi-gas of electrons ($Z = Z_1^*$ or $Z_2^*$)
($\Gamma \equiv Z^2e^2/a_i kT$ ; $r_S \equiv a_e/a_B$ ; $\theta \equiv kT/\varepsilon_F \equiv 4/(9\pi)^{1/3}(n_e \Lambda^3_e)^{2/3}$ ; $\Lambda^2_e \equiv 2\pi\hbar^2/m_e kT$ ; $a_j^3 \equiv 4\pi n_j/3$)

| | $Z$ | $T_C$, a.u | $(n_e)_C$,cc$^{-1}$ | $P_C$, a.u. | $\Gamma_C$ | $(r_S)_C$ | $(n_e \Lambda^3_e)_C$ | $(\theta)_C$ |
|---|---|---|---|---|---|---|---|---|
| $Z = Z_1^*$ | 34.6 | 6.38. | 2.24 $10^{25}$ | 11.4 | 140 | 0.416 | 3.30 | 2.91 |
| $Z = Z_2^*$ | 45.4 | 9.29 | 3.96 $10^{25}$ | 28.4 | 181 | 0.344 | 3.26 | 2.89 |

## CRITICAL EXPONENTS

Remarkable feature of two discussed pseudo-critical points at $Z = Z_1^*$or $Z = Z_2^*$ is the non-standard values of all critical exponents in comparison with the ordinary (van der Waals like) critical exponents that correspond to the case of OCP{c} with the charge number $Z$ beyond the discussed interval $Z_1^* \div Z_2^*$. For example, at the latter case ($Z < Z_1^*$ or $Z > Z_2^*$), the standard density$\leftrightarrow$temperature relation is valid

$$(\rho - \rho_C) \sim |T - T_C|^{1/2}$$

For the pseudo-critical points ($Z = Z_1^*$ or $Z = Z_2^*$) the following relation may be proved:

$$(\rho - \rho_C) \sim |T - T_C|^{1/3}$$

Direct calculation gives:

$$|\rho/\rho_C - 1| \cong 4.57\, |T/T_C - 1|^{1/2} \qquad (Z = 1)$$
$$|\rho/\rho_C - 1| \cong 4.07\, |T/T_C - 1|^{1/3} \qquad (Z = Z_1^* \cong 34.6)$$

## SATURATION CURVE

Similar violation is observed for saturation ($P_{St} \leftrightarrow T_{St}$) curve. So-called Plank – Gibbs rule (equal slope of saturation curve at $T = T_C - \varepsilon$ and critical isohore at $T = T_C + \varepsilon$) is valid for an ordinary critical point ($Z < Z_1^*$ or $Z > Z_2^*$),

$$(\mathrm{d}P/\mathrm{d}T)_{St} = (\partial P/\partial T)_{V_c}$$

It is not evident (see Figure 3), but it can be proved that this rule is *not valid* for pseudo-critical points ($Z = Z_1^*$ or $Z = Z_2^*$).

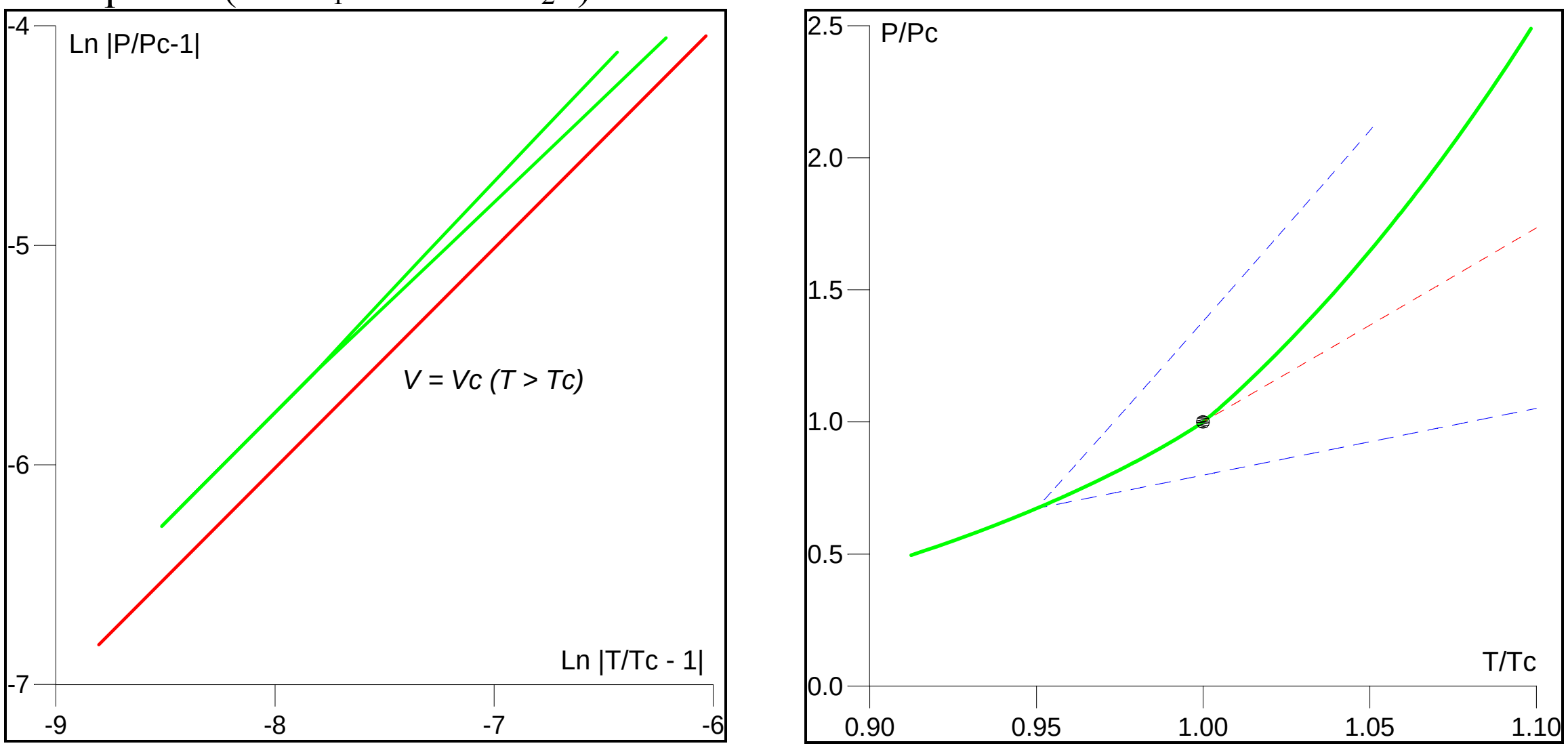

**Figure 3.** Saturation curve and isochors in reduced coordinates for the Single OCP{c} of classical point charges with uniform compressible background of ideal fermi-gas of electrons at boundary value of charge number $Z = Z_l^* \cong 34.6$. Notations: *1*- sublimation; *2* - melting; *3* - pseudo-critical «termination» point; *4* - critical isochore; *5,6* - sub-critical isochors.

**Figure 4.** The same as on Figure 3 in Log↔Log - coordinates. Notations: *1-4* - as on Figure 1.

This statement is illustrated on Figure 4. Small deviation in position of binodal (curves 1,2) and critical isochore (curve 4) corresponds to the small difference in slope of both the curves at pseudo-critical point (Figure 3).